\documentclass[10pt,twocolumn,letterpaper]{article}

\usepackage{iccv}

\makeatletter
\@namedef{ver@everyshi.sty}{}
\makeatother
\usepackage[accsupp]{axessibility}
\usepackage{times}
\usepackage{epsfig}
\usepackage{graphicx}
\usepackage{amsmath}
\usepackage{amssymb}

\usepackage[utf8]{inputenc}
\usepackage{mathtools}
\usepackage{booktabs}
\DeclareMathOperator*{\argmax}{argmax}
\DeclareMathOperator*{\argmin}{argmin} 
\usepackage{amsmath,amssymb}
\usepackage[flushleft]{threeparttable}
\usepackage{array,booktabs,makecell}
\usepackage[font=small]{caption}
\usepackage[usenames, dvipsnames]{color}
\usepackage{amsfonts}
\usepackage{amsmath}
\usepackage{tabularx,booktabs}
\usepackage{threeparttable}
\DeclareMathOperator{\E}{\mathbb{E}}
\usepackage{tikz}
\newcommand*\circled[1]{\tikz[baseline=(char.base)]{
    \node[shape=circle,draw,inner sep=0.5pt] (char) {#1};}}

\usepackage[pagebackref=true,breaklinks=true,letterpaper=true,colorlinks,bookmarks=false]{hyperref}

\iccvfinalcopy 

\def\iccvPaperID{8} 

\ificcvfinal\fi

\begin{document}

\title{A Dual Adversarial Calibration Framework for Automatic Fetal Brain Biometry}

\author{Yuan Gao$^{1}$\thanks{Equal contribution.}\ \  \thanks{Corresponding author: Yuan.Gao2@eng.ox.ac.uk}
~ ~ Lok Hin Lee$^{1}$\footnotemark[1]
~ ~ Richard Droste$^{1,2}$\footnotemark[1]
~ ~ Rachel Craik$^{1,3}$ 
~ ~ Sridevi Beriwal$^1$\\
~ ~ Aris Papageorghiou$^1$
~ ~ Alison Noble$^1$ \\
\normalsize $^1$ University of Oxford ~ ~ $^2$ Amazon ~ ~ $^3$ Kings College London 
}

\maketitle
\ificcvfinal\thispagestyle{empty}\fi

\begin{abstract}
This paper presents a novel approach to automatic fetal brain biometry motivated by needs in low- and medium- income countries. Specifically, we leverage high-end (HE) ultrasound images to build a biometry solution for low-cost (LC) point-of-care ultrasound images. We propose a novel unsupervised domain adaptation approach to train deep models to be invariant to significant image distribution shift between the image types. Our proposed method, which employs a \textbf{Dual Adversarial Calibration} (DAC) framework, consists of adversarial pathways which enforce model invariance to; i) adversarial perturbations in the feature space derived from LC images, and ii) appearance domain discrepancy. Our \textbf{Dual Adversarial Calibration} method estimates transcerebellar diameter and head circumference on images from low-cost ultrasound devices with a mean absolute error (MAE) of 2.43mm and 1.65mm, compared with 7.28 mm and 5.65 mm respectively for SOTA.
\end{abstract}

\section{Introduction}

Pregnancy dating is a crucial part of obstetric care because antenatal care and interventions aimed at improving pregnancy outcome rely on knowledge of the gestational age (GA). Measurements of the size of specific fetal head anatomies is routinely performed to estimate and validate GA. Biometries used include the fetal skull Head Circumference (HC) and the Transcerebellar Diameter (TCD) \cite{tcd}, easily measured on ultrasound images from high-end (HE) machines. US images from HE imaging machines have high imaging contrast, high imaging definition and low speckle noise compared to low-cost (LC) ultrasound images which are acquired with point-of-care (POC) ultrasound probes with greater varied image appearance and hence quality \cite{dietrich2017point}. However, HE imaging may not be available in resource-constrained areas. 

Previous literature has considered automated approaches for fetal brain biometry on HE 2D ultrasound (US) images for GA estimation. \cite{2d1} use contour detection and graph cuts for HC estimation. \cite{2d2} propose a regional convlutional neural network for detection of key anatomical structures. \cite{sobhaninia2019fetal} use U-Nets for HC segmentation and measurement. More recently, \cite{zhang2020direct} directly regresses HC measurements from ultrasound images without segmentation and \cite{lee2020calibrated} directly regresses GA from fetal head images using a Bayesian neural network. However, these methods estimate on mid-end US images, which may not be available in resource-constrained settings. 

In this paper, we consider jointly learning HC and TCD automated fetal biometry from partially labelled US images acquired with a HE ultrasound machine (GE Voluson E8) combined with unlabelled data from a LC POC ultrasound probe (Konted C10R) for biometry on LC images. This is clinically relevant in a case where interobserver variation on ground truth labelling on LC images is high, due to the reduced image quality and fuzzy edges, and can be useful where a central corpus of well-labelled HE images are available for use for validation and inference on LC images. The core assumption is that domain invariant representations for feature extraction can be jointly learned from LC and HE ultrasound images, and unsupervised learning can calibrate the model in the LC domain so to produce consistent predictions. To this end, we propose a \textbf{Dual Adversarial Calibration} (DAC) approach exploiting two adversarial pathways. One pathway forces predictions from LC and HE US images to lie on the same output manifold by training a segmentation network with a discriminator which learns to classify between them. The other pathway forces the LC output to be invariant to self-paced adversarial noise perturbations. Additionally, we propose a novel asymmetric domain augmentation technique specifically designed to cope with the appearance discrepancy between HE and LC images. Experimental results presented show that our proposed approach significantly improves the performance of HC and TCD biometry on LC US images compared to a neural network trained on HE images alone. Ablation experiments also reveal that our dual adversarial pathways lead to a network that is able to learn useful feature representations that are domain invariant.
\begin{figure*}
  \includegraphics[width=\textwidth]{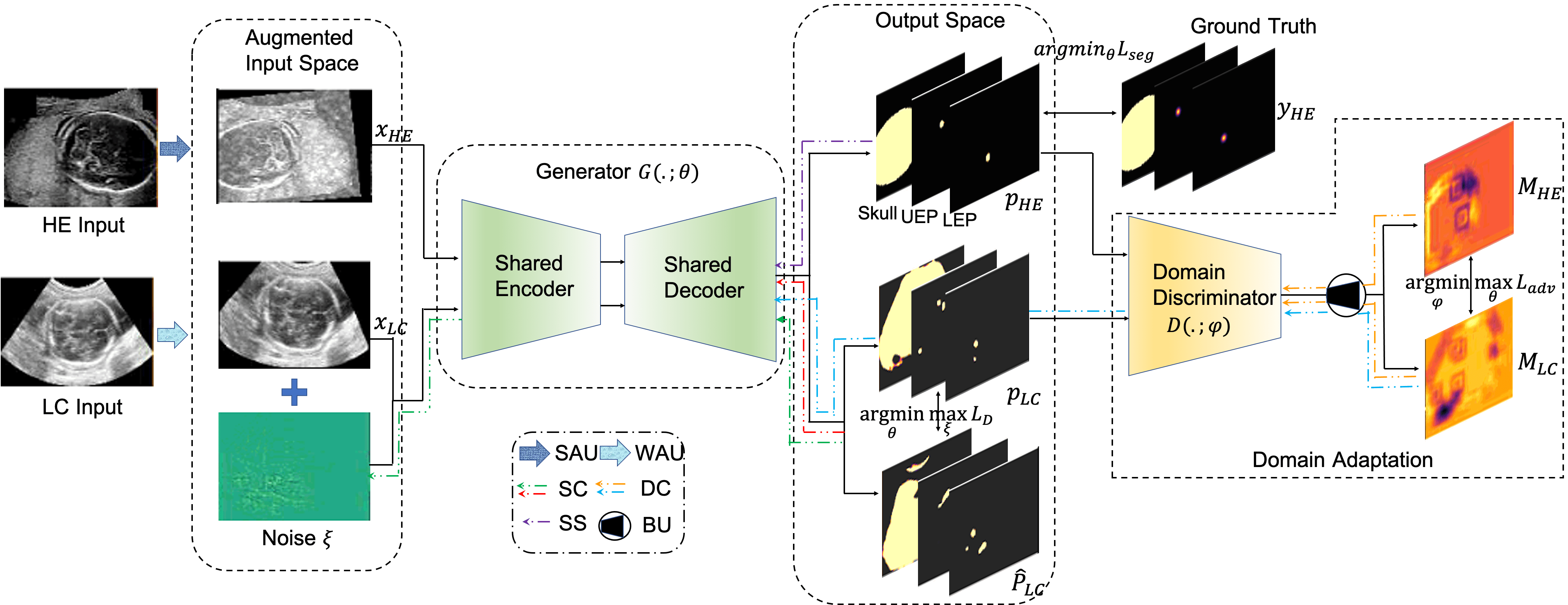}
  \caption{Overview of the training procedure for our proposed \textbf{dual adversarial calibration} network, which is trained with two adversarial signals simultaneously. 1) a Domain Calibration (DC) pathway designed to learn domain-invariant segmentation, and 2) a Segmentation Calibration (SC) pathway designed to learn robust segmentation on the unlabelled LC images. SAU: Strong Augmentation; WAU: Weak Augmentation; SS: Supervised Segmentation (on HE images); BU: Bilinear Upsampling. UEP, LEP explained in section \ref{lep}.}
  \label{pics:method}
\end{figure*}
\subsection{Contributions.}
Our contributions are three-fold:
1) we leverage the supervised learned knowledge from HE ultrasound images to significantly improve biometry estimation on LC ultrasound images; 2) we introduce a novel dual adversarial unsupervised intra-modality domain and semantic transfer for this purpose; 3) our results are shown to be competitive compared to SOTA for both automated HC and TCD estimation on LC US images.


\section{Related Work}
\subsection{Medical Cross-Modality Domain Adaptation.}
Medical cross-modality domain adaptation aims to retain network performance from the distribution change from an image resulting from one imaging modality to another. Examples of imaging modalities include magnetic resonance imaging (MRI), computed tomography (CT) and ultrasound (US) imaging. Prior literature focused on cross-modality domain adaptation between CT/MR \cite{ct1,ct2,ct3,ct4}. In detail, \cite{ct1} uses an adversarial domain adaptation module to map target input features to the output domain space. \cite{ct2} use a tumour-preserving cycle-consistency loss to map CT and MRI images before training a U-Net to segment lung MRI scans. \cite{ct3} use a shared encoder space with adversarial based domain adaptations for the segmentation of cardiac structures. \cite{ct4} share convolutional kernels between MRI and CT images, but use modality specific normalization layers to improve cross-modality performance. Comparatively, literature on cross-modality domain adaptation involving US is limited. \cite{mu} generate synthetic MR fetal head images from US scans, but validation is limited as only appearance is evaluated without segmentation results.


\subsection{Intra-Modality US Domain Adaptation}
Previous literature on intra-modality US domain adaptation focused on adaptation between different HE imaging devices \cite{degel2018domain,liu2021generalize,uda3}. \cite{degel2018domain} use a u-net for left atrium segmentation in 3D ultrasound, and incorporate the imaging device used as prior knowledge during inference. \cite{liu2021generalize} use a hierarchical style transfer network to modify image appearances from a target domain to the source domain for fetal head and abdomen segmentation. \cite{uda3} considered intra-domain ultrasound adaptation for image classification using mutual information minimization. However, in all the examples above, the domain gap in considered was limited as both source and target domain images were acquired with HE ultrasound machines with similar imaging capabilities.

\subsection{Low-Cost Ultrasound Probe Image Analysis}
Three recent papers investigate learning from low-cost POC probes \cite{2d4,van2019automated,2d3}. Gao et al.\cite{2d4} developed an image quality assessment framework to identify frames from LC US that can be used for manual TCD measurement by a sonographer. Van den Heuvel et al.\cite{van2019automated} investigate gestational age estimation on downsampled HE images to evaluate GA estimation on degraded images, but do not directly evaluate from LC POC probes. Maraci et al.\cite{2d3} segments the cerebellum for TCD measurement, but the LC US image led to poor segmentation performance. 

\section{Method}
\begin{figure}
    \centering
\includegraphics[width=\linewidth,height=3in]{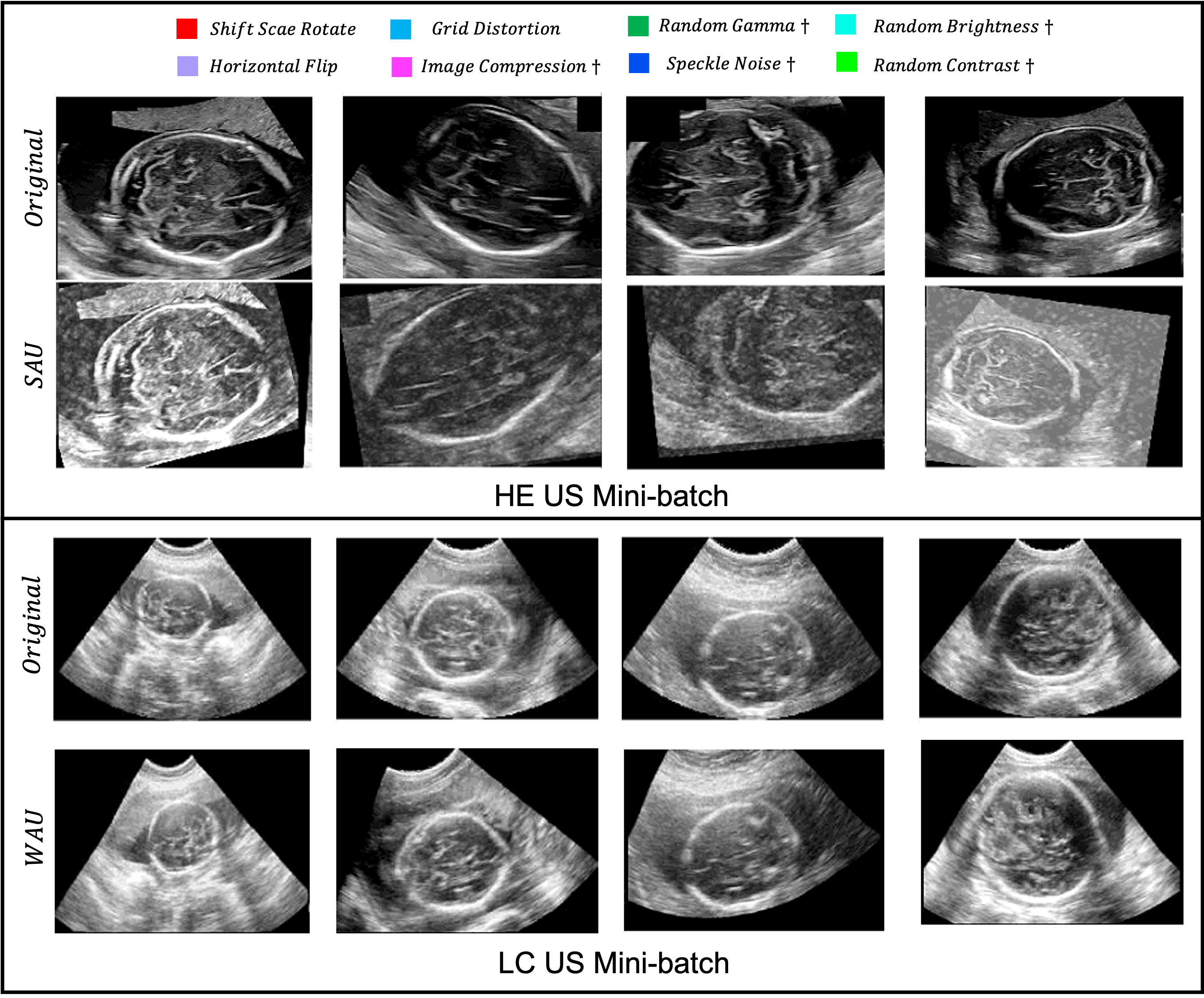} 
\caption{Asymmetrical Domain Augmentation. SAU: Strong Augmentation; WAU: Weak Augmentation. $\dag$: Extra augmentations included in SAU.}
\label{fig:aug}
\end{figure}
In this section, we present the \textbf{Dual Adversarial Calibration} framework in detail, which is illustrated in Fig.\ref{pics:method}. The framework consists of four main parts: 1) a domain dependent asymmetrical augmentation module of the input space; 2) a segmentation network consisting of a shared encoder-decoder framework; 3) a domain calibration (DC) adversarial pathway for semantic transfer on the predicted output space; 4) a segmentation calibration (SC) cycle pathway for unsupervised semantic transfer.  
\subsection{Asymmetrical Domain Augmentation.}

We observe that in practice, HE data are fairly consistent in imaging quality and appearance, whereas LC data can vary quite substantially. Key anatomies, such as the cerebellum and the thalamus are always clearly visible on HE data, but may not be so on LC data, as imaged structures do not have clear edges and acoustic artefacts such as shadows and speckle can lead to further image degradation. We therefore augment each input domain asymmetrically. Specifically, weak augmentation is applied on LC data to maximize network generalization whilst preserving the spatial visibility of key anatomies by the inclusion of linear and non-linear grid distortions, and horizontal flipping. Strong augmentation is applied on HE data to simulate noisy images including random gamma, random brightness and contrast adjustment, image compression and artificial speckle noise. We qualitatively observe these data augmentations act to decrease domain gap between HE and LC data, as shown in Figure \ref{fig:aug}, which leads to a stronger response to domain calibration.

\subsection{Core Segmentation Network.}
\label{lep}
We formulate fetal biometry via segmentation tasks. The HC is derived from segmentation of the fetal skull. However in LC US images, segmentation of the cerebellum is challenging due to low image contrast and noise from signal attenuation. We therefore adopt a strategy inspired by TCD measurement in real-clinical practice, where two points are placed on the cerebellum boundary edge for TC diameter estimation. Thus, the network predicts the upper extreme point (UEP) and lower extreme point (LEP) of the cerebellum, from which the TCD is estimated. 

We therefore target two image analysis tasks - fetal skull segmentation for the HC, and UEP and LEP detection for TCD estimation. We employ a U-Net based segmentation network, represented by $\{G(x; \theta), x \in (x_{LC}, x_{HE})\}$ where $x_{LC}$ and $x_{HE}$ are examples of HE and LC input US images respectively and $\theta$ represents model parameters. We use ResNet18 \cite{res} as an encoder-decoder backbone with residual blocks (fine to coarse: {64, 64, 128, 256 ,512} feature channels) for the encoder and residual blocks with 2D bilinear upsampling (coarse to fine: {256, 128, 64, 32, 16} feature channels) for the decoder with skip connections between the encoder-decoder. Our segmentation head consists of a single $3\times3$ 2D convolutional layer to map the decoded feature maps (16 channels) to 3 channels, corresponding to the skull, UEP and LEP predictions. We then apply a pixel-wise sigmoid function to the segmentation output to obtain a pixel-wise probability map for each anatomy. To address class imbalance, we use a DICE loss to train the segmentation model defined as:
\begin{figure}
    \centering
\includegraphics[width=\linewidth,height=1.5in]{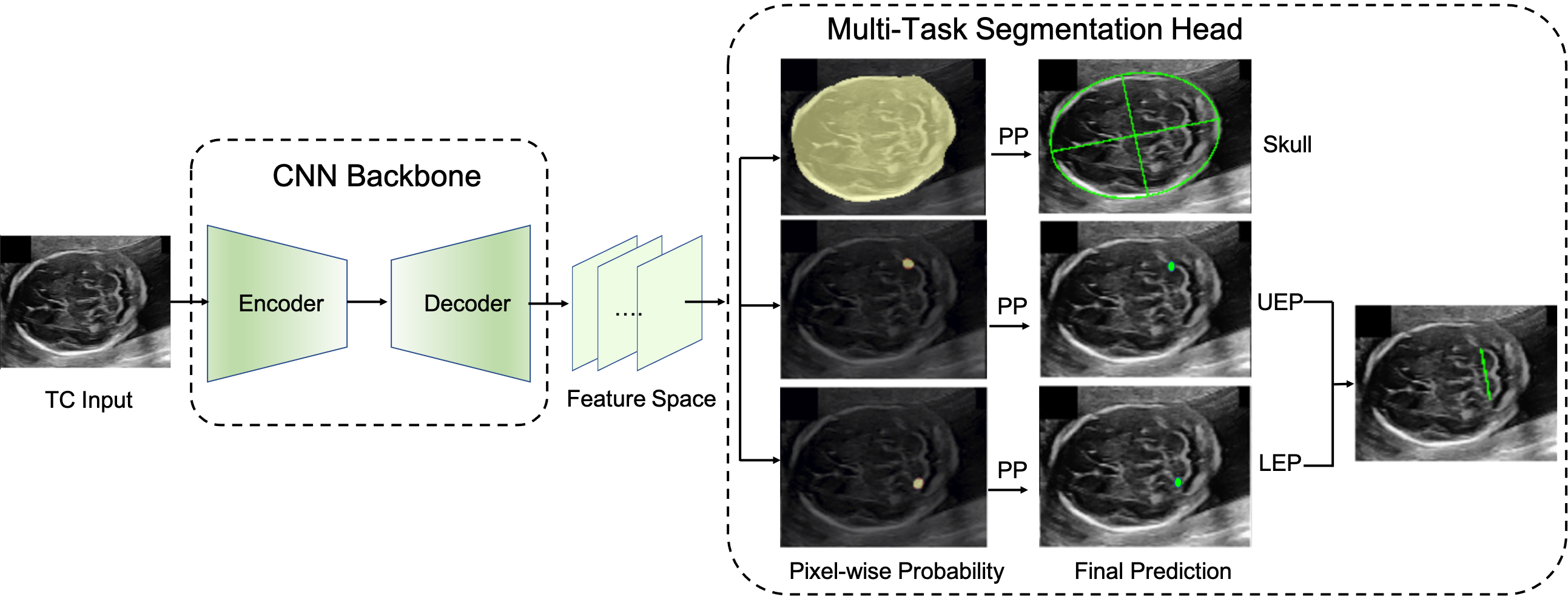} 
\caption{The pipeline from segmentation to the final biometry.}
\label{fig:seg}
\end{figure}
\begin{equation}
L_{seg}= 1-\frac{2\sum_{h,w,c}y_{HE}p_{HE}}{\sum_{h,w,c}y^{2}_{HE}+\sum_{h,w,c}p^{2}_{HE}}
\label{seg}
\end{equation}
where $y_{HE}\epsilon \mathbb{R}^{h\times w\times c}$ and $p_{HE}\epsilon \mathbb{R}^{h\times w\times c}$ are the ground truth annotations and the pixel-wise probability maps for HE images, h, w and c are height, width and number of classes respectively (c=3 in our experiments).

The whole pipeline from input to final biometric estimation is depicted in Figure \ref{fig:seg}. We got the segmentation probability maps for each structure i.e. Skull, UEP and LEP. For computing HC, we segment the entire fetal head, retrieve a skull contour from the probability map, then fit an ellipse to the contour, obtain the center, major and minor axis, and rotation of the fitted ellipse. For computing TCD, we perform non-maximum suppression to find the pixel with greatest probability for UEP and LEP, then draw a line between the two estimated point locations.

\subsection{Domain Calibration Pathway.}
We observe that $x_{HE}$ and $x_{LC}$ have very different intensity distributions. This reflects in imaging quality discrepancy, and networks trained on one do not perform well on the other. To address this, we introduce an adversarial pathway to calibrate the underlying output space to be invariant to the input domain \cite{output}. Specifically, the output space can be modelled as a low-dimensional manifold that contains simple representations and rich semantic information about target anatomies. By minimizing the distance between HE US predictions ${p}_{HE} = G(x_{HE}; \theta)$ and LC US predictions ${p}_{LC} = G(x_{LC}; \theta)$, the model can learn specific target anatomical regions of interest from a low dimensional representational space. We adopt a Least-Square GAN \cite{lsgan} loss for domain adaptation. The domain adaptation is modelled with the objective: 

\begin{equation}
\begin{split}
\argmin_{\varphi}L_{adv}(D) & := \frac{1}{2} \E_{p \sim {p}_{out}(HE)}[(D({p}_{HE};\varphi)-1)^2]\\ &  + \frac{1}{2} \E_{p \sim {p}_{out}(LC)}[(D(p_{LC};\varphi))^2]
\end{split}
\label{eq1}
\end{equation}

\begin{equation}
\argmin_{\theta}L_{adv}(G) := \frac{1}{2} \E_{ p \sim {p}_{out}(LC)}[(D({p}_{LC};\varphi)-1)^2]
\label{eq2}
\end{equation}

Here $D(p; \phi)$ represents the input probality map ${p}$ to our discriminator parameterised by $\phi$. The discriminator consists of five convolutional layers (fine-to-coarse: 64, 128, 256, 512, 1 feature channels) with a kernel size of 3 and stride of 1 and each followed by a leaky ReLU parameterised by 0.2 and a max pooling layer. 

\begin{figure*}[t]
  \includegraphics[width=\textwidth,height=3in]{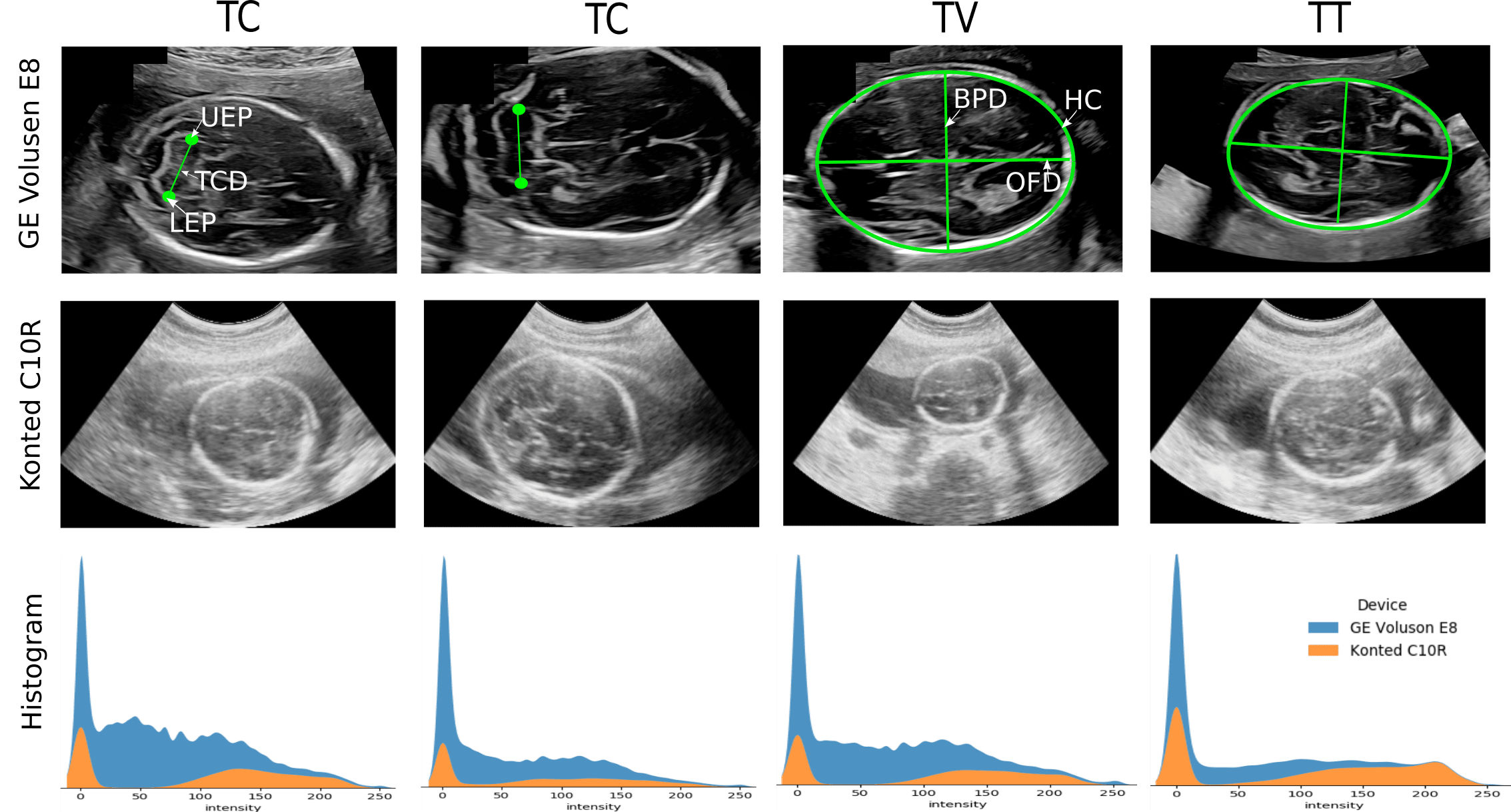}
  \caption{Examples of fetal brain biometry planes acquired by different devices and pixel intensity distributions. HE scanner: GE Voluson E8, LC probe: Konted C10R. TC: Transcerebellar planes, TV: Transventricular planes, TT: Transthalamic planes, UEP: Upper Extreme Point, LEP: Lower Extreme Point, TCD: Transcerebellar Diameter, HC: Head Circumference, BPD: Biparietal Diameter, and OFD: Occipito-Frontal Diameter.}
  \label{pics:intro}
\end{figure*}

\subsection{Segmentation Calibration Pathway.}
The generator $G(x, \theta)$ is trained on supervised segmentation with the labelled source input $x_{HE}$ with available ground truth $y_{HE}$. We further regularize the model using unlabelled $x_{LC}$ images by using adaptive perturbations by noting that the predicted segmentation should be locally smooth to adversarial perturbations in the input $x_{LC}$ images \cite{vat}. As shown in Fig.\ref{pics:method}, we generate an augmented LC input tuple $X = \{(x_{LC}, x_{LC}+\xi), \xi \sim N(0, 1)\}$ and prediction tuple $P = (p_{LC}, \hat{p}_{LC})$ from the segmentation network $G(X; \theta)$. $\xi$ is self-paced and updated with a cycle pathway by maximizing the distance between $(p_{LC}, \hat{p}_{LC})$, generating $\xi_{adv}$, an adversarial perturbation, then minimising the Kullbeck-Leibler loss $KL\left [G(x;\theta) \parallel G(x+\xi_{adv}) \right ]$ to reduce network sensitivity to noise perturbations. To do this we compute the derivative of $L_{D}$ w.r.t. $\xi_{adv}$ defined as $g_{\xi}=\bigtriangledown_{\xi_{adv}}L_{D}$ evaluated with backpropagation, and the perturbation as $\xi_{adv}=\epsilon \frac{g_{\xi}}{\left \| g_{\xi} \right \|_{2}}$, where $\epsilon$ is a hyper-parameter that controls the strength of perturbation. The goal is therefore for the network to be robust to the perturbations so as to produce consistent outputs from the perturbed inputs. The distance loss $L_{D}$ is therefore defined as:

\begin{equation}
\begin{split}
&\argmin_{\theta} L_{D}(x,\xi,\theta):= \\&\E_{x\sim p_{data}(LC)} KL[ [G(x_{LC};\theta)]\parallel [G(x_{LC}+\xi_{adv};\theta)]] \\&\text{s.t.} \ \xi_{adv}:= \argmax_{\xi} \left \{ L_{D}(x_{LC},\xi,\theta);\left \| \xi \right \|_{2}\leq \epsilon  \right \}
\end{split}
\end{equation}

\subsection{Optimisation.}
Our model optimization is performed in a two-step process during model training. The domain discriminator is first optimized by minimizing the loss $L_{adv}(D)$ while the generator is frozen. The generator is then subsequently optimized by optimizing the joint loss $L_{joint}$, defined as:

\begin{equation}
\centering
\begin{split}
&L_{joint}(x_{HE},x_{LC},\xi, \theta)= \\&L_{seg}(X_{HE},\theta)+ \alpha L_{D}(X_{LC},\xi,\theta) + \beta L_{adv}(G)
\label{obj}
\end{split}
\end{equation}
where the hyperparameters $\alpha$ and $\beta$ determine the strength of segmentation and domain calibration respectively.

\section{Experiments and Results}


\begin{table*}

\caption{Dataset description for training and evaluation of our model. Planes are split between test and training on a subject basis to prevent data leakage.}

\resizebox{\textwidth}{!}{%
\begin{tabular}{lccccc}
\hline
\multicolumn{1}{l|}{}           & \multicolumn{1}{c|}{No. of Subjects} & \multicolumn{1}{c|}{Acquisition Device} & \multicolumn{1}{c|}{Training (Seg. Labels)}   & \multicolumn{2}{c|}{Testing (Biometry {[}mm{]})} \\ \cline{4-6} 
\multicolumn{1}{l|}{}           & \multicolumn{1}{c|}{}                & \multicolumn{1}{c|}{}                   & \multicolumn{1}{c|}{TC Planes}                & TC Planes (HC)             & TT/TV Planes (TCD)         \\ \hline
\multicolumn{1}{l|}{HE Dataset} & \multicolumn{1}{c|}{540}             & \multicolumn{1}{c|}{GE Volusen E8}      & \multicolumn{1}{c|}{519 (Labelled HC \& TCD)} & -                          & -                          \\
\multicolumn{1}{l|}{LE Dataset} & \multicolumn{1}{c|}{560}             & \multicolumn{1}{c|}{Konted GEN1 C10R}   & \multicolumn{1}{c|}{418 (Unlabelled)}         & 387 (Labelled)             & 526 (Labelled)             \\ \hline
                                & \multicolumn{1}{l}{}                 & \multicolumn{1}{l}{}                    & \multicolumn{1}{l}{}                          & \multicolumn{1}{l}{}       & \multicolumn{1}{l}{}      
\end{tabular}%
}
\label{tab:gae-est}
\end{table*}

\subsection{Datasets.}We have two different datasets acquired from two different clinical studies, examples of which can be seen in Fig. \ref{pics:intro}. HE and LC images were acquired from a clinical US scanner and a POC probe respectively described in Table \ref{tab:gae-est}, along with the distribution of scans used for training and testing. The difference in image quality is apparent in Fig. \ref{pics:intro}. This includes reduced imaging contrast between tissues of high echogenicity (fetal skull) and low echogenicity (internal brain tissue), reduced clarity of internal brain structures such as the cerebellum and ventricles, reduced edge sharpness as can be seen from the boundary of the skull. All of these features are clinically used for determination of TCD and HC. Furthermore, image resolution is reduced and there is increased noise in LC images. We used TC planes from the HE dataset segmented with HC and TCD measurement points along with unlabelled TC planes from the LC dataset to train our model. The performance of our model was then evaluated by comparison between biometry from inferred segmentations and the biometry extracted by an expert sonographer. 

Images from each source differed in input resolution (784$\times$1008 px HE, 228$\times$378 px LC). All images were resized to 448$\times$576px. Ground truth pixel maps for the fetal skull were labelled $x_i \in \{0, 1\}$ for pixels inside and outside the skull respectively. For TCD estimation, instead of labelling the entire cerebellum, a Gaussian kernel was centered on the sonographer annotated key points i.e. UEP and LEP used as the ground truth for training. 
As seen in table \ref{tab:results}, we achieve mean TCD error of 2.43 mm and mean HC error of 1.65mm on the best performing model.

\subsection{Network Training.}

Our model was implemented with pytorch 1.4.0 and trained on a single Quadro RTX 5000 GPU. $G(\theta)$ was optimized with Nesterov accelerated SGD with an initial learning rate of 0.1, momentum of 0.9 and a weight decay of $10^{-3}$. $D(\phi)$ was optimized with Adam with an initial learning rate of $10^{-4}$ and weight decay of $10^{-4}$. Both $D(\phi)$ and $G(\theta)$ were optimized for 70 epochs with a minibatch size of 4. Both learning rates were multiplied by a factor of 0.1 after epochs 40 and 60. Grid search was performed on a logarithmic scale for $\alpha$ and $\beta$; $\alpha = 10^{-1}$ and $\beta =10^{-3}$ gave the best performance.
\begin{table*}[t]
\centering
\caption{Ablation study of our method and comparison to SOTA for HC and TCD estimation on LC test dataset. Mean Absolute Error (MAE) $\pm$ std is reported.}
\label{tab:results}

\resizebox{\textwidth}{!}{%
\begin{tabular}{l|c|c|c|c|c}
\midrule\midrule
Method                              & Aug.                              & DC, Adpt. Loss, Adpt. Space                   & SC                  & \begin{tabular}[c]{@{}c@{}}TCD \\ (mean $\pm$ SD [mm])\end{tabular} & \begin{tabular}[c]{@{}c@{}}HC\\  (mean $\pm$ SD [mm])\end{tabular} \\ \hline
\circled{1} W/o        & \checkmark, Weak                  &       -                                       &      -               &   46.63$\pm$8.64                                                                    &   36.27$\pm$7.81                                                                 \\
\circled{2} W/o         & \checkmark, Strong                &        -                                      &       -              &    30.96$\pm$7.46                                                                 &       22.52$\pm$7.54                                                             \\
\circled{3} DC          & \checkmark, Asymmetrical          & \checkmark, V-GAN Loss, out. space           &      -               &     13.80$\pm$3.91                                                                &   10.21$\pm$3.63                                                                 \\
\circled{4} DC          & \checkmark, Asymmetrical          & \checkmark, LS-GAN Loss, feat. space         &   -                  & 8.64$\pm$1.42                                                       & 6.53$\pm$1.11                                                      \\
\circled{5} DC          & \checkmark, Asymmetrical          & \checkmark, LS-GAN Loss, out. space          &       -              & 7.93$\pm$1.64                                                       & 4.31$\pm$1.25                                                      \\
\circled{6} DAC         & \checkmark, Asymmetrical          & \checkmark, LS-GAN Loss, feat. space         & \checkmark          & 4.62$\pm$0.46                                                       & 3.26$\pm$0.43                                                      \\
\textbf{\circled{7} DAC} & \textbf{\checkmark, Asymmetrical} & \textbf{\checkmark, LS-GAN Loss, out. space} & \textbf{\checkmark} & \textbf{2.43$\pm$0.37}                                              & \textbf{1.65$\pm$0.31}                                             \\ \midrule\midrule
CycleGAN \cite{cycle}               & \checkmark, Asymmetrical                                 & \checkmark, Cycle-GAN Loss, in. space               & -                & 14.79$\pm$3.20                                                      & 9.69$\pm$3.19                                                     \\
CyCADA \cite{caca}                  & \checkmark, Asymmetrical                                & \checkmark, CyCADA Loss, in.$\&$feat. space     & -                   & \textbf{7.28$\pm$2.72}                                                       & \textbf{5.65$\pm$2.25}                                                    \\
UCMDA \cite{ct1}      & \checkmark, Asymmetrical                                & \checkmark, Adversarial loss, in feat. space                                          & -                   & 8.13$\pm$2.30                                                       & 6.74$\pm$2.19                                                      \\
SIFA \cite{uda2}       & \checkmark, Asymmetrical                                 & \checkmark, SIFA Loss, in.$\&$feat.$\&$out. space                                            & -                   & 8.69$\pm$1.21                                                       & 6.32$\pm$1.04    \\             
\midrule\midrule
\end{tabular}%
}

\begin{tablenotes}
\item{*} Note: W/o: Without Adaptation; DC: Domain Calibration; SC: Segmentation Calibration; DAC: Dual Adversarial Calibration.
\end{tablenotes}
\end{table*}
\subsection{Results and Ablation Study}

\begin{figure}
    \centering
\includegraphics[width=\linewidth,height=3.5in]{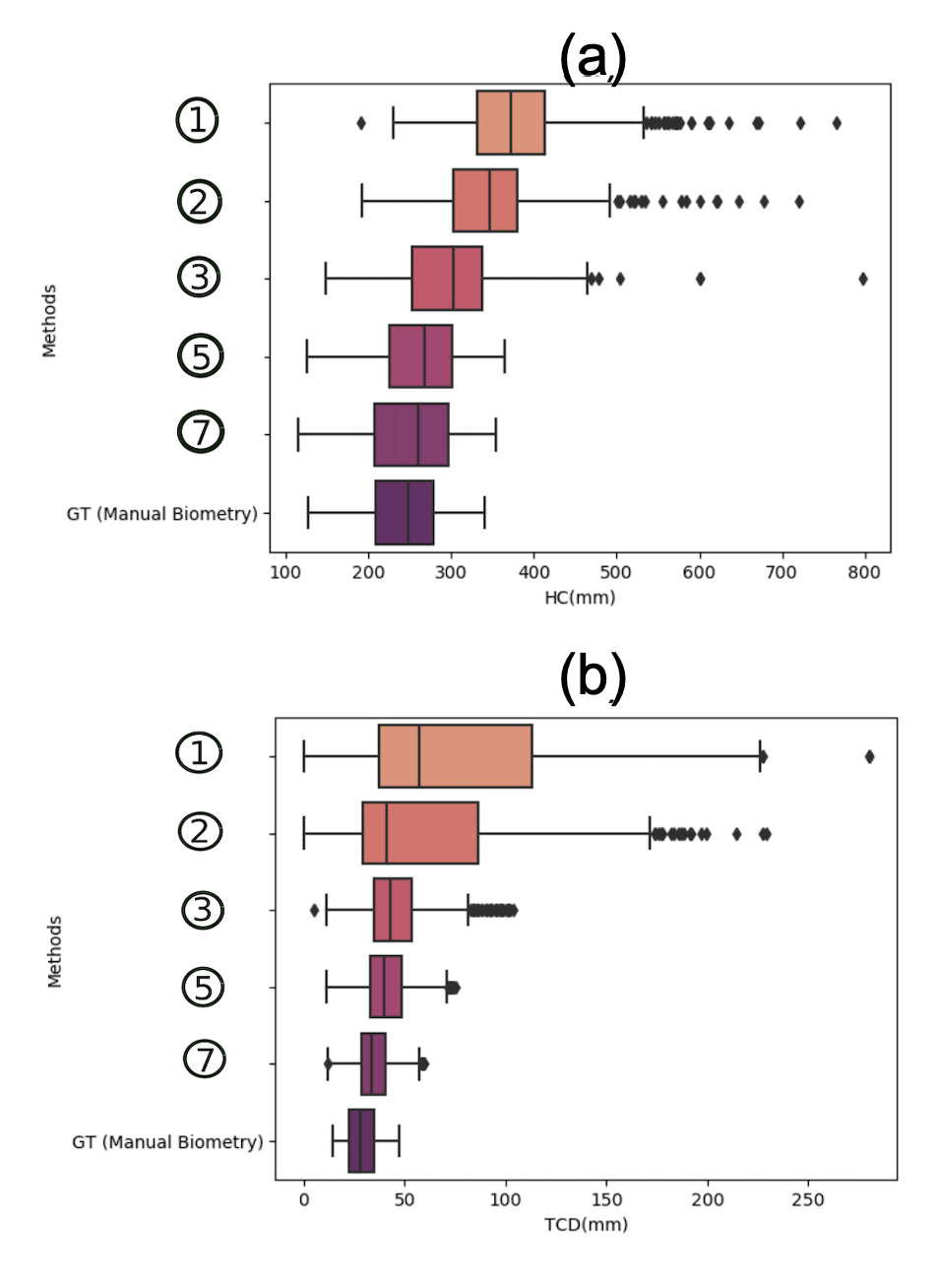} 
\caption{Biometry distributions on LC test dataset from different models numbered 1-7 as in Table \ref{tab:results}. (a) predicted HCs (b) predicted TCDs for different models. GT: expert manual biometry.}
\label{fig:ds}
\end{figure}

\begin{figure*}[t]
  \includegraphics[width=\textwidth]{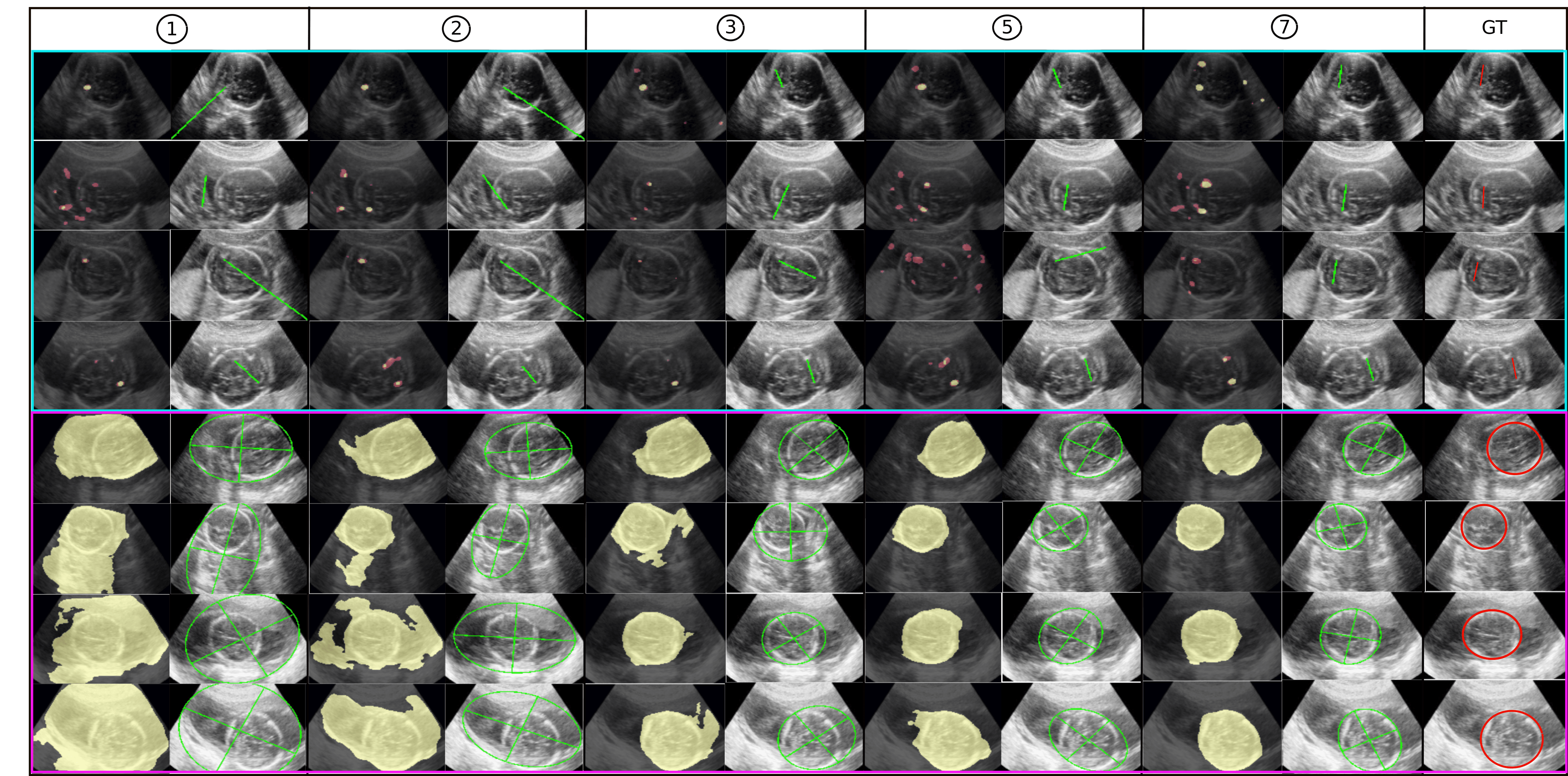}
  \caption{Example outputs from different models numbered 1-7 as in Table \ref{tab:results}, evaluated on LC test images (Blue box: TCD, purple box: HC). The GT column represents the ground truth.}
  \label{pics:r1}
\end{figure*}
To better understand the individual contributions from each component of our model, we perform an ablation study which composed of three settings: 1) we train the segmentation network with labelled HE images and the directly apply to LC images at test time; 2) we investigate domain adaptation from either feature or output space by applying adversarial training; 3) in addition to domain adaptation, we incorporate the self-paced unsupervised branch for segmentation calibration. We show in Table \ref{tab:results} (\circled{1}, \circled{2}) which are models trained with HE images directly evaluated on LC images. Direct estimation of biometries from LC images from a model trained with HE images performs poorly as expected due to the substantial domain shift. Inclusion of adversarial calibration (\circled{3}, \circled{4}, \circled{5}) significantly improves biometry estimation. We also find that training with a LS-GAN adversarial loss outperforms a Vanilla GAN loss, and adversarial calibration on the output space leads to better localization compare to feature space (\circled{5}, \circled{7} vs.\circled{4}, \circled{6}), which suggests that the output manifold is a suitable rich environment for domain calibration. Further addition of adversarial segmentation calibration (\circled{6}, \circled{7}) leads to the best performance with MAE of 1.65mm for HC and 2.43mm for TCD estimation.

Also, as can be seen in Fig. \ref{fig:ds} (a) and (b), without domain adaptation i.e. \circled{1}, \circled{2}, there are a number of outliers and the biomertry is significantly out of the distribution compared to the expert's manual biometry for both the HC and TCD. By incorporating the adversarial training \circled{3}, \circled{5}, the model's predicted distribution is shifted towards the ground truth. DAC \circled{7} results in the most closed distribution to the expert biomerty. Additionally, we found HC predicted by DAC model tends to be better agreed with expert measurements, compared to TCD and the HC mean absolute error is smaller than TCD's. This may because the appearance of the TCD landmark is prone to be affected by image quality and artefacts, however, the skull signal is bright and consistent, which is less affected by the imaging quality changes. 

\subsection{Comparison with SOTA.}
We compare our proposed model with current SOTA in the lower rows of Table \ref{tab:results}. We found that domain adaptation methods that adapt from the input space (CycleGAN \cite{cycle}) give a high estimation error, as only style is preserved by internal image content and anatomies are not well conserved, especially for TC images. Taking into account the feature space (CyCADA \cite{caca}) leads to better performance, but continues to misjudge points of the cerebellum. We also investigated the SOTA methods for unsupervised cross-modality domain adaptation, UCMDA and SIFA \cite{ct1,uda2}. UCMDA\cite{ct1} proposed a domain critic module (DCM) that minimizes difference between domains in the feature space. SIFA uses a shared encoder and performs domain adaptation in the output space, and use cycle-GAN based identity loss to learn domain invariant features. We find that these cross modality models not necessarily works on our intra-modality adaptation task and underperform compared to our best performing model \circled{7}. This may because intra-modality domain shift is relatively moderate compared to cross-modality and the cross-modality models are too complex to fit on the moderate domain shift. Our proposed DAC design focus on aligning the intra-modality domain shift by simply incorporating a self-paced distribution alignment process. The self-paced calibration helps to align the intra-domain feature space and make models more robust against artefacts and variation in imaging quality, introduced by LC probes.  

\subsection{Qualitative Results.}

As shown in Fig. \ref{pics:r1}, without domain or segmentation adversarial calibration, models (\circled{1}, \circled{2}) fail to localize at least one of the TCD measurement points and fail to segment the skull for HC. We find that including domain calibration (\circled{3}, \circled{5}) leads to noisy probability maps for the TCD measurement points, but the final prediction (after non-maximum suppression) become more accurate. Using a LS-GAN loss leads to smoother adversarial calibration for domain adaptation and which leads to more localized prediction of the TCD measurement points (\circled{5} vs. \circled{3}). In addition to domain calibration, we find that segmentation calibration (reflected in Table\ref{tab:results} \circled{5} vs. \circled{7}) increases the performance of TCD estimation more than HC estimation. This suggests that adversarial segmentation perturbation helps the network to localize of small targets in a noisy environment. Our complete model \circled{7} outperforms all of the above. It correctly identifies the TCD measurement points and segments the fetal skull, even in challenging examples.  
 
\section{Conclusion}
This paper addresses the problem of domain adaptation from clinical high-end ultrasound images to low cost point-of-care ultrasound images with greater varied imaging quality and increased noise. We proposed a novel dual adversarial calibration framework which enables the network to learn invariant features to both image types, and leverages high-end ultrasound images to enable a solution for accurate automatic biometry on LC US images. Our approach outperforms current SOTA for domain calibration and semi-supervised learning methods applied to this task.

\subsubsection*{Acknowledgements}
We acknowledge the ERC (ERC-ADG-2015 694 project PULSE), the EPSRC (EP/R013853/1, EP/T028572/1) and the MRC (MR/P027938/1).

{\small
\bibliographystyle{ieee_fullname}
\bibliography{bib}
}

\end{document}